\documentclass[final,3p,times]{elsarticle}
\biboptions{sort&compress}

\usepackage{hyperref}

\usepackage{graphicx}

\usepackage{amssymb}
\usepackage{amsmath}
\usepackage{upgreek}
\usepackage{lineno}
\usepackage{gensymb}
\usepackage[utf8]{inputenc}
\usepackage[dvipsnames]{xcolor}

\usepackage{fancyhdr}
\usepackage[normalem]{ulem}

\journal{}

\begin{document}

\color{red} \textbf{This is the accepted version of the article published open access in Ultramicroscopy: \newline \href{}{https://doi.org/10.1016/j.ultramic.2024.114071  } \newline} \color{black}

\begin{frontmatter}

\title{Toward quantitative thermoelectric characterization of (nano)materials by \textit{in-situ} transmission electron microscopy}

\author[label1,label2]{Simon Hettler\corref{cor1}}
\author[label1,label2]{Mohammad Furqan}
\author[label1]{Andrés Sotelo}
\author[label1,label2,label3]{Raul Arenal\corref{cor1}}
\address[label1]{Instituto de Nanociencia y Materiales de Aragón (INMA), Universidad de Zaragoza, Zaragoza, Spain}
\address[label2]{Laboratorio de Microscopías Avanzadas (LMA), Universidad de Zaragoza, Zaragoza, Spain}
\address[label3]{ARAID Foundation, Zaragoza, Spain}
\cortext[cor1]{hettler@unizar.es,arenal@unizar.es}

\begin{abstract}

We explore the possibility to perform an \textit{in-situ} transmission electron microscopy (TEM) thermoelectric characterization of materials. A differential heating element on a custom \textit{in-situ} TEM microchip allows to generate a temperature gradient across the studied materials, which are simultaneously measured electrically. A thermovoltage was induced in all studied devices, whose sign corresponds to the sign of the Seebeck coefficient of the tested materials. The results indicate that \textit{in-situ} thermoelectric TEM studies can help to profoundly understand fundamental aspects of thermoelectricity, which is exemplary demonstrated by tracking the thermovoltage during \textit{in-situ} crystallization of an amorphous Ge thin film. We propose an improved \textit{in-situ} TEM microchip design, which should facilitate a full quantitative measurement of the induced temperature gradient, the electrical and thermal conductivities, as well as the Seebeck coefficient. The benefit of the \textit{in-situ} approach is the possibility to directly correlate the thermoelectric properties with the structure and chemical composition of the entire studied device down to the atomic level, including grain boundaries, dopants or crystal defects, and to trace its dynamic evolution upon heating or during the application of electrical currents.

\end{abstract}

\end{frontmatter}

\section{Introduction}

Thermoelectric materials are an important building block for generation of green energy from heat and a huge range of different materials are being studied \cite{Merrill.2015,Twaha.2016,Jaziri.2020,Yan.2022}. In general, the thermoelectric effect is a complex phenomenon and the Seebeck coefficient strongly depends on crystallinity, presence of dopants, contaminants or impurities in the material \cite{Nadtochiy.2019,Jacob.2021,Torres.2022}. In the development of thermoelectric bulk materials, different strategies are followed to improve the figure of merit, $ZT$, including grain-boundary engineering \cite{Jacob.2021} or doping \cite{Torres.2022}. Nanomaterials, nano-composites or nano-structured devices are also promising candidates for thermoelectric materials \cite{Small.2003,Fukuhara.2018,Nadtochiy.2019}. The miniaturization of devices, the use of nanomaterials and the complex interplay between the structure/composition of materials and their thermoelectric properties have led to numerous different approaches for electric and thermoelectric characterization of (nano)materials \cite{Qi.2013,Liu.2016,Hu.2020,Bishara.2021}. 

Transmission electron microscopy (TEM), combined with related spectroscopic techniques, allows the in-depth analysis of structure and composition down to the atomic scale \cite{Pennycook.2011,Deepak.2015}. Technological developments have improved the energy resolution in electron energy-loss spectroscopy (EELS) and allow the local study of phonons by scanning (S)TEM \cite{Krivanek.2014,Yan.2021}, which are important for heat transport phenomena. Recently, \textit{in-situ} TEM studies have gained significant importance as they allow to study the evolution of materials under external stimuli. One branch of \textit{in-situ} experiments are electrical studies, where the influence of an applied voltage \cite{Zhang.2017,Sato.2017,Nukala.2021} or current \cite{Hettler.2021} on a material is investigated. Here, we explore the possibility to measure the Seebeck coefficient and thermoelectric properties of materials, including individual nanomaterials, by \textit{in-situ} TEM. The combination of thermoelectric characterization and TEM is promising for gaining a deeper understanding of fundamental relationships in thermoelectric materials as, for example, the role of grain boundaries.

\section{Materials and Methods}

\subsection{Device setup}

\begin{figure}[h]
    \centering
    \includegraphics[width=0.9\linewidth]{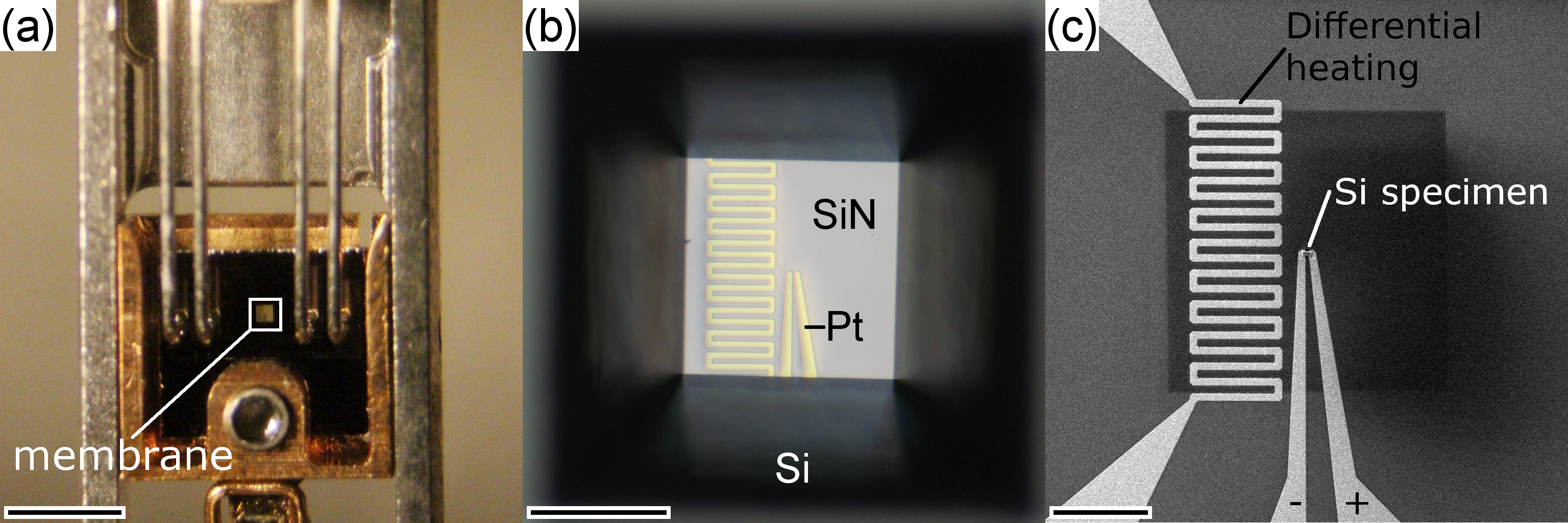}
    \caption{(a) Photograph of the tip of the \textit{in-situ} TEM holder with thermoelectric microchip inserted. The position of the membrane has been indicated. (b) Optical microscopy image from the membrane with heating device and contact pads taken from the back side of the \textit{in-situ} TEM microchip. (c) SEM image taken from the front side of the Si device. Polarity of the I-V measurements of the devices is indicated. Scale bars are (a) 2~mm, (b) 200~$\upmu$m and (c) 100~$\upmu$m. }
    \label{F:Fig1}
\end{figure}

The general concept to measure the Seebeck coefficient is to generate a temperature gradient along the material to be studied and simultaneously measure the generated electrical potential. To combine such a measurement with TEM investigations, a microchip that fits in an \textit{in-situ} TEM sample holder was designed and fabricated by microelectromechanical systems (MEMS) technologies. The chip fabrication procedure was described elsewhere in detail \cite{Hettler.2024_prep}. Figure~\ref{F:Fig1} shows a photograph, an optical microscopy image and a scanning electron microscopy (SEM) image of one of these chips, which consist of a differential heating device and two contact pads (10~nm Ti + 150~nm Pt) on top of a free-standing low-stress silicon nitride membrane (thickness 1~$\upmu$m). The differential heating element is located on one side of the membrane in order that an applied heating current ($I_H$) would create a temperature gradient along the specimen placed between the contact pads (Figure~\ref{F:Fig1}c), which are located in the center of the membrane. To measure the voltage induced in thermoelectric materials by a specific temperature gradient, I-V curves are acquired, which provide higher accuracy compared to a simple voltage measurement and give additional information on the resistivity.  The polarity of the voltage applied to the specimen is indicated in Figure~\ref{F:Fig1}c.

\subsection{Investigated materials}
\label{MM.materials}
To prove that reliable thermoelectric measurements are possible with the proposed \textit{in-situ} approach and to illustrate the advantage of the approach, we studied five materials with different thermoelectric properties by preparing devices using the fabricated chips. We chose p-doped (boron) single-crystalline silicon and a single-crystalline molybdenum (source: sputter target, structural analysis in supplementary information (SI), section SI1) as reference standards for a semiconducting and metallic material, respectively. As a third well-studied reference material, strontium-doped calcium cobaltite (CCO: Ca\textsubscript{2.93}Sr\textsubscript{0.07}Co\textsubscript{4}O\textsubscript{9}) misfit-layered compound (MLC) was used. The CCO bulk material was prepared as described in \cite{Torres.2022} by attrition milling (2h at 600 rpm) starting from SrCO\textsubscript{3}, CaCO\textsubscript{3} and CoO. The mixture was calcinated at 850~\degree C during 1~h, cold uniaxially pressed into 25 mm diameter disks under 250 MPa, and subjected to hot-uniaxial pressing at 900~\degree C and 51 MPa for 1 h. For this CCO material, two devices were fabricated with the aim to study the relationship between different crystal directions of the asymmetric material and its thermoelectric properties, as discussed below. 

An example for a material with negative Seebeck coefficient are nanotubes (NTs) based on the MLC (LaS)\textsubscript{1.14}-TaS\textsubscript{2} \cite{Hettler.2020,Merrill.2015}. Finally, two devices were prepared based on amorphous Ge (aGe) and amorphous C (aC) thin films to perform dynamic \textit{in-situ} experiments (see section~\ref{S:Dyn}). aGe and aC/aGe/aC thin films were prepared by electron-beam evaporation (Auto 500 Boc Edwards) from Ge pellets (99.999\%, Lesker) and highly ordered pyrolitic graphite (HOPG) sheets, respectively. Base vacuum was 5~10\textsuperscript{-7}~mbar, which increased to 1~10\textsuperscript{-5}~mbar during the evaporation processes. The aC/aGe/aC thin film was prepared without venting the chamber.  The films were deposited onto freshly cleaved mica sheets and then transferred to Cu and SiN TEM grids in a water-based floating process for further inspection and \textit{in-situ} device preparation.

\subsection{In-situ device preparation}

In order to study these materials, a piece of each has to be deposited between the two contact pads on the \textit{in-situ} TEM chip, which was achieved with a dual-beam instrument (Thermo Fisher Scientific Helios 650) with SEM and focused ion beam (FIB). For bulk materials, FIB-based preparation methods for \textit{in-situ} studies have been published \cite{Duchamp.2014} and their first step is the preparation of a conventional TEM lamella: a vertical slice of material with 1~$\upmu$m thickness, 12~$\upmu$m width and a depth of approximately 5~$\upmu$m is taken out from the bulk material and transferred to a TEM copper grid. In a second step, the slice is flipped and transferred to the contact pads of the \textit{in-situ} chip, where a hole has been milled previously between the pads by FIB. The lamella can be thinned prior to transfer to the chip and/or directly on the chip by using a sample holder with inclined surface. Here, we did a combination of both. At first we thinned the sheet on the Cu grid down to 1~$\upmu$m thickness over the entire width and polished the surface that will be in contact to the chip to guarantee a good contact. Except for the CCO $\parallel$ (see below) device, which was transferred without addtional pre-thinning, we further thinned the central part of the sheet from one side down to 700~nm forming a kind of bridge. This shape facilitates the thinning after the transfer of the flipped sheet to the chip as it moves the incidence of the focused ion beam  away from the surface. The final thinning of the central part and the removal of the Pt cover from the TEM lamella preparation was done using an inclined sample holder that allows milling with the FIB parallel to the chips' surface. SI2 describes this preparation exemplary for the CCO $\perp$ (see below) device. 

For the MLC NT and the aGe thin-films, the preparation needs to be different and we employed a recently developed support-based method \cite{Hettler.2024_prep}. SI3 shows this preparation process exemplary for the MLC NT device, which consists in first preparing a conventional TEM sample by drop-casting the nanomaterial on a holey silicon nitride chip and the selection of a suitable NT by TEM. Using FIB, the NT and the support are transferred to the \textit{in-situ} chip, where finally the silicon nitride is milled away to leave the NT as unique bridge between the contacts.

\subsection{Electron microscopy}

The MEMS chip is designed to fit in a DENSsolutions Wildfire holder with 4 pins for a ThermoFisherScientific/FEI instrument, which allows TEM analysis and electrical characterization of the chip at the same time. TEM, SAED and scanning (S)TEM analysis have been conducted in two aberration-corrected microscopes (Thermo Fisher Scientific Titan\textsuperscript{3} (image-corrected) and Titan low-base (probe-corrected)) operated at 300 keV electron energy. HAADF STEM imaging was done with a convergence and collection angle of 25~mrad and 48~mrad, respectively. Energy-dispersive X-ray spectroscopy (EDX) was performed in STEM mode with an Ultim X-MaxN 100TLE detector (Oxford Instruments). Electron energy-loss spectroscopy (EELS) data was acquired in STEM mode with a collection angle of 29.5~mrad using a Gatan Image Filter Tridiem 865 spectrometer.

  During the electrical measurements, the specimen was irradiated with a relatively large parallel beam (TEM mode). An effect of the electron beam on the thermoelectric characterization could not be observed under these conditions. However, the device setup could allow to perform electron-beam induced current (EBIC) measurements or to investigate possible electron-beam heating effects. 

\subsection{Raman spectroscopy}

A confocal Raman Alpha 300 M+ (WiTec) was used for acquisition of Raman spectra, with a 633 nm laser operated at 0.25 mW power and a 50x objective. The spectrometer was operated with 600 grooves/mm grating.

\subsection{Electrical characterization}

Thermoelectric measurements were performed with two Keithley instruments devices: A Keithley Instruments 2450 SourceMeter (Tektronix) was employed for I-V characterization and a 2611A System Source Meter for simultaneously applying the heating current to the differential heating device. 

\section{Results and discussion}

\begin{figure}[t]
    \centering
    \includegraphics[width=0.9\linewidth]{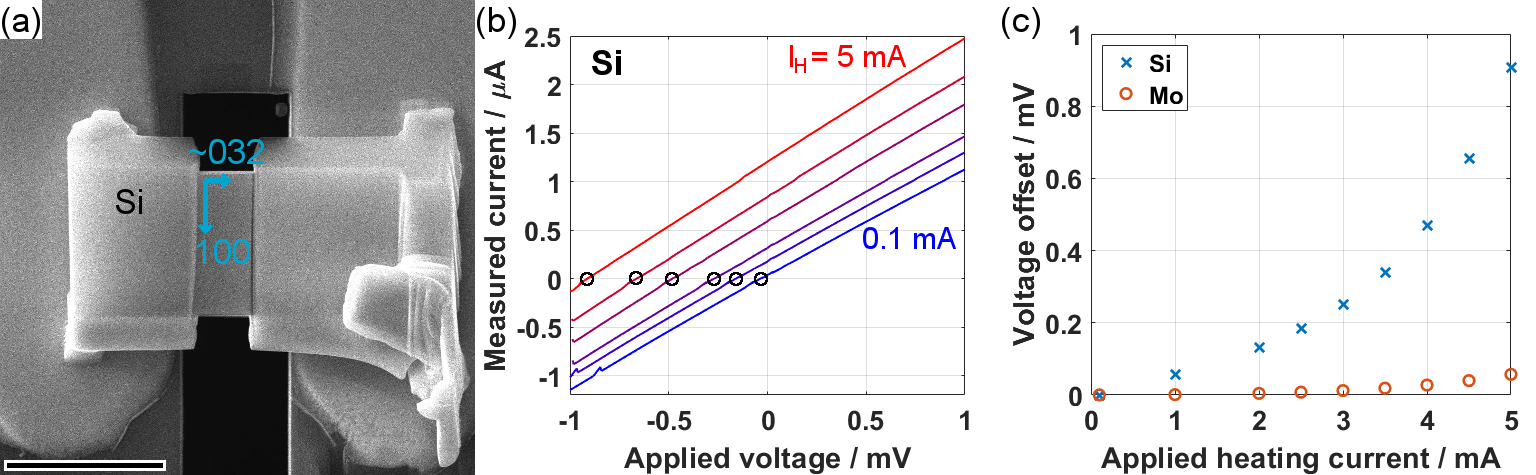}
    \caption{(a) SEM image of the fabricated device showing the Si sheet deposited between the contact pads of the \textit{in-situ} chip. Orientation of the Si crystal lattice is indicated. Scale bar is 5~$\upmu$m. (b) I-V characterization of the device for $I_H$~=~0.1 (blue), 2, 3, 4, 4.5 and 5~mA (red). (c) Plot of voltage offset depending on $I_H$ for the Si device shown in (a) and a similar device prepared from single-crystalline Mo. }
    \label{F:Fig2}
\end{figure}

\subsection{Reference materials and device characterization}

Figure~\ref{F:Fig2}a shows a SEM image of the prepared p-doped Si device, where the material was cut into a cuboid with defined geometry in the central area above the hole in the membrane. The crystal orientation is 100 in vertical direction and approximately 032 in direction of the current flow between the contacts (SAED analysis in SI4). Energy-dispersive X-Ray spectroscopy (EDX) indicates a minor surface oxidation and Ga incorporation (SI5). Figure~\ref{F:Fig2}b depicts an I-V characterization of the device shown in (a) obtained at different applied heating currents $I_H$ between 0.1 (blue curve) and 5~mA (red curve).

With increasing $I_H$ applied to the differential heating device, the I-V curve shifts along the x-axis and the gradient slightly increases. The gradient change is linked to a slight decrease in average resistance from 890~$\Upomega$ to 775~$\Upomega$ due to an increase in temperature and a related improvement of the semiconducting Si and also likely due to a reduction of contact resistance. The shift can be attributed to the generation of a voltage induced by a temperature gradient along the sample. To measure this shift, a linear regression $I(V)=-V/R+I_{0}$ is performed on the I-V curves and the shift is calculated from $V_0=-\frac{I_{0}}{R}$ and is marked by black circles in Figure~\ref{F:Fig2}b. The negative sign is due to the employed polarity (Figure~\ref{F:Fig1}c), as the temperature gradient is opposite to the applied voltage. The voltage offset is determined by the value of $V_0(I_H)$ relative to $V_0$($I_H$=0) to account for possible minor intrinsic currents. $V_0$ is shown in Figure~\ref{F:Fig2}c in dependence of $I_H$ for the Si device and a similar sample prepared from single-crystalline Mo. Both curves show a clear increase of the voltage offset with $I_H$, which corresponds with the expected increase of the temperature gradient along the material. The voltage offset is positive, which agrees with the (positive) sign of the Seebeck coefficient of both materials \cite{Fulkerson.1968,Fiflis.2013}. A quantitative assessment of the measured voltages is discussed below. 

To check a possible contribution to the thermolectrical measurement by the chip itself, the raw \textit{in-situ} chip, without deposited material, was characterized in the same way. No measurable contribution was observed (SI6). 

\subsection{Devices made of misfit-layered compounds}

\begin{figure}[h]
    \centering
    \includegraphics[width=0.85\linewidth]{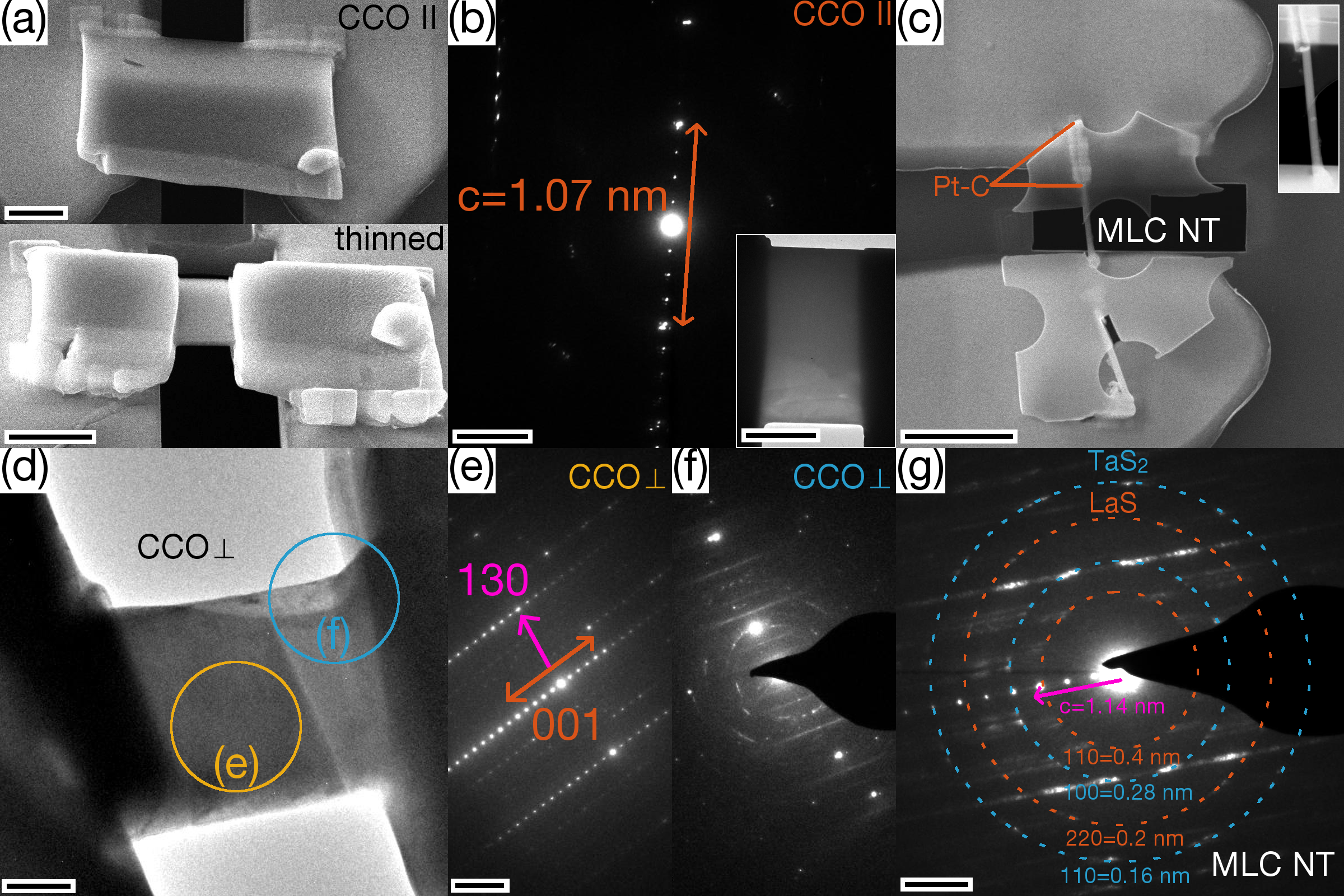}
    \caption{ Analysis of the devices based on MLC materials. (a) Two SEM images of the unthinned (top) and thinned (bottom)  CCO $\parallel$ device, respectively. (b) SAED pattern acquired from the thinned CCO sheet (inset TEM image) showing the c-axis periodicity of the MLC stack. (d) TEM image of the thin part of the CCO $\perp$ device with two grains visible. (e,f) SAED patterns acquired as marked in (d). (c) SEM image and inset low-magnification STEM image (height 4~$\upmu$m) of the MLC-NT device. (g) SAED pattern evidencing its nanotubular structure. Scale bars are (a) 2~$\upmu$m, (b) 3~nm\textsuperscript{-1}, inset: 1~$\upmu$m, (c) 4~$\upmu$m, (d) 60~nm, (e,f) 4~nm\textsuperscript{-1} and (g) 1~nm\textsuperscript{-1}.}
    \label{F:Fig3}
\end{figure}

Figure~\ref{F:Fig3} shows the structural analysis of the three devices made of misfit-layered compounds (MLCs). MLCs consist of two different layered materials stacked alternatively upon each other in c crystal direction, leading to a strong asymmetry between in-plane and out-of-plane directions of the crystal structure \cite{Merrill.2015}. Two devices were prepared from the bulk CCO material with different crystal orientations with respect to the induced temperature gradient. Two SEM images of the first CCO device before and after FIB thinning for TEM studies are shown in Figure~\ref{F:Fig3}a. The microstructure of the Sr-doped CCO bulk material was analyzed using a standard TEM lamella (SI7), which revealed that the material possesses a micro-crystalline structure with most grains exhibiting a similar direction of the c-axis. The c-axis corresponds to the direction of the applied pressure during the hot-pressing process, see section~\ref{MM.materials}. Grain sizes are small in c-direction (approximately 500~nm) and large in a-b directions (several $\upmu$m). A few grains and grain boundaries spanning horizontally across the device can be seen in the inset TEM image in Figure~\ref{F:Fig3}b, which also depicts a selected-area electron diffraction (SAED) pattern obtained from the thin part of the device. The SAED pattern confirms that the c-axis is perpendicular to the grain boundaries visible in Figure~\ref{F:Fig3}b, implying that for this device, the electrical current flow is predominantly in-plane and thus parallel to the layers of the MLC. This device is denoted as CCO $\parallel$ in the following. The c-axis periodicity of the CCO material is measured to be 1.07 nm from the SAED pattern, which agrees with previously reported values on bulk CCO \cite{Miyazaki.2002}.
\medskip

\medskip

Figure~\ref{F:Fig3}d-f details a second CCO device, denoted as CCO $\perp$. Figure~\ref{F:Fig3}d shows a SEM image of the device, which was prepared such that the current flow is roughly perpendicular to the layered structure (parallel to the c-axis) and that a single grain boundary falls in the thinned central part. Figure~\ref{F:Fig3}d shows a bright-field (BF) TEM image of this device after orientation of the left grain into a high-order zone axis (TEM stage at $\upalpha$=23\degree, $\upbeta$=23\degree). The image clearly reveals the presence of two grains with the left grain oriented in zone-axis, appearing darker compared to the right grain. Figure~\ref{F:Fig3}e and \ref{F:Fig3}f show two SAED patterns obtained from the left and right part of the device as marked in Figure~\ref{F:Fig3}d, corresponding to the two grains. The patterns confirm the orientation of the left grain in a higher-order zone-axis. An analysis of the reflections indicates the presence of [001] (1.07~nm) and [130] (0.15~nm) reflections, corresponding to a [-310] zone-axis orientation. Note that due to the monoclinic structure of the CCO material, [001] and [130] are not perpendicular to each other. On the other hand, the right grain does not exhibit a higher order orientation of the crystal with respect to the electron beam at this particular stage tilt (Figure~\ref{F:Fig3}f). The analysis of the device tilted to an angle where the right grain is oriented in [010] zone-axis (TEM stage at $\upalpha$=16\degree, $\upbeta$=19\degree) is shown in SI8. As the 001 direction roughly corresponds to the $\upbeta$ tilt direction, the angle between the c-axis directions of the two grains is estimated by the difference in $\upalpha$ tilt, which roughly corresponds to 7\degree. An EDX spectrum of this CCO device is shown in SI5 and confirms the presence of all elements of the CCO device, including the strontium doping.

The third MLC device consists in a NT made of (LaS)\textsubscript{1.14}-TaS\textsubscript{2}. A SEM image of the device is shown in Figure~\ref{F:Fig3}c. The silicon nitride membrane utilized for the sample transfer to the chip was milled away in the central region in order that the NT forms the only conductive connection between the contact pads (see preparation in SI3). A rather long Pt-C focused ion beam induced deposition (FIBID) was needed to contact the NT to the upper contact pad. A SAED pattern acquired from the central part of the NT (Figure~\ref{F:Fig3}g) reveals that the complex structure of the MLC NT was preserved during the preparation \cite{Hettler.2020}: Reflections are seen from the two different subsystems LaS (orange dashed rings) and TaS\textsubscript{2} (blue dashed rings) as well as from the c-axis periodicity of the stack (pink arrow), which is perpendicular to the NT axis, as seen in Figure~\ref{F:Fig3}c. The tubular structure is recognized from the streaking reflections caused by the bending of the layers. The damage-free transfer with minimum Pt-C contamination is also seen from a high-angle annular dark field (HAADF) STEM image acquired from the edge of the NT (SI3) and an EDX spectrum of the MLC-NT device (SI5).

\begin{figure}[t]
    \centering
    \includegraphics[width=0.9\linewidth]{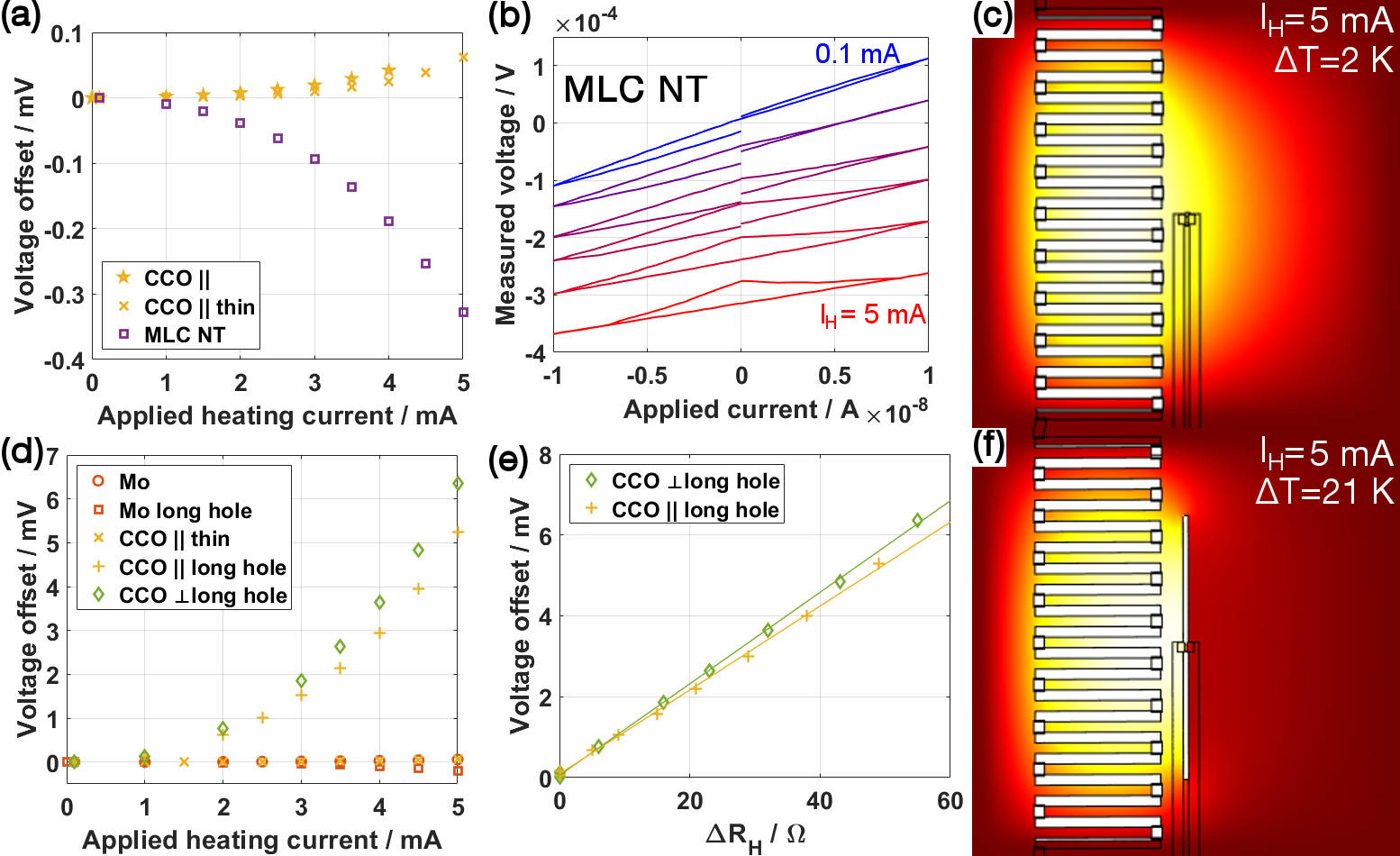}
    \caption{(a) Voltage offset for the CCO $\parallel$ and the MLC NT devices shown in Figure~\ref{F:Fig3}a-c and h-j. respectively. (b) I-V curves for the MLC NT for $I_H$~=~0.1 (blue), 2.5, 3.5, 4, 4.5 and 5~mA (red) showing a hysteresis loop. (d) Comparison of the voltage offset for Mo and CCO $\parallel$ devices with short and long hole together with CCO $\perp$ device with long hole. (e) Voltage offset over resistance increase $\Delta R$ of the heating device with linear fit.  (c,f) Maps of the temperature distribution within the silicon nitride membrane obtained from COMSOL simulations assuming a (c) short and (f) long hole, respectively. White color corresponds to the maximum temperature (317 K (c) and 324 K (f)) and dark red to room temperature.}
    \label{F:Fig4}
\end{figure}

The CCO $\parallel$ and NT devices were analyzed (thermo)electrically similar to the Mo and Si devices (Figure~\ref{F:Fig2}) and the determined voltage offset $V_0$ in dependence of $I_H$ is shown in Figure~\ref{F:Fig4}a. The CCO $\parallel$ device was measured twice: Firstly without thinning and still including the Pt protection cover from the TEM lamella fabrication (yellow pentagrams in Figure~\ref{F:Fig4}a, top image in Figure~\ref{F:Fig3}a) and secondly after FIB thinning (yellow crosses in Figure~\ref{F:Fig4}a, bottom image in Figure~\ref{F:Fig3}a). Only a small difference is visible between the two curves indicating that the impact on the thermoelectric effect of the Pt protection cover is small, even though it strongly contributes to the electrical conductivity of the device (see following section). Also, the thickness of the lamella (1~$\upmu$m thinned down to 350~nm in the central part) only plays a minor role as the Seebeck coefficient is a parameter mainly intrinsic to the material, which does not dependent on smaller modifications of its dimensions, and if these do not affect the mixture of grains in the material. Nevertheless, the observed minor decrease of the measured voltage offset after thinning is not expected in the simple picture of an additive combination of the thermoelectric properties in a composite material. Under such an assumption, a device made of the combination of the CCO material (154~$\upmu$V/K \cite{Torres.2022}) and the Pt cover (-5~$\upmu$V/K) should show a lower voltage than the thinned CCO material alone. However, previous studies have shown that a combination of a thermoelectric material based on cobalt oxide with a conductive metal can actually increase the Seebeck coefficient, which explains the observed trend \cite{Wang.2008,Sotelo.2013}. The magnitude of the voltage offset for the CCO $\parallel$ is comparable to the Mo device, which is surprisingly low, as the Seebeck coefficient of CCO bulk material (154~$\upmu$V/K) is much higher than the one of Mo \cite{Fiflis.2013} (discussion below).

The measured voltage offset for the NT device is plotted by purple squares in Figure~\ref{F:Fig4}a. In contrast to the previous devices, the NT device is the only one that shows a negative voltage offset, which agrees with previous studies showing a negative Seebeck coefficient for chalcogen-based MLC materials \cite{Merrill.2015}. Figure~\ref{F:Fig4}b shows the I-V curves for the NT device for different values of $I_H$. A hysteresis effect can be observed in all of the curves, which we attribute to a space-charge region formed in the long Pt-C FIBID contact. Such memristive behaviour has been frequently observed in nanoscale devices \cite{Strukov.2008}. Nonetheless, the effect of the differential heating device is obvious, clearly shifting the curves in opposite direction as for the Si device. In this MLC nanotube device, only the lower, straight part of the hysteresis loop has been used to determine the voltage offset.

\subsection{Electrical properties}

In addition to the thermoelectric characterization, the chips also facilitate an \textit{in-situ} TEM measurement of the electrical properties of the materials. The electrical conductivity $\upsigma$ can be obtained from the measured resistance and the sample geometry. The data obtained from the devices at room temperature ($I_H$=0) are summarized in SI9.

As the measurements were conducted in a two-probe setup, the contribution of contact resistance needs to be considered, which was approximately 135~$\Upomega$ for the employed chips. For semiconducting specimens with dimensions typical for TEM investigations, the contact resistance is only a small part of the total device resistance. For example, the two thinned CCO MLC devices show resistances of 10~k$\Upomega$ (CCO $\parallel$) and 80~k$\Upomega$ (CCO $\perp$), which lead to conductivities of 71~Sm\textsuperscript{-1} and 54~Sm\textsuperscript{-1}, respectively. These values are one order of magnitude higher when comparing to bulk measurements \cite{Torres.2022}. This could be attributed to the fact that pores, absent in the studied microscopic devices but frequent in the bulk (see SI7), significantly contribute to the resistivity of the CCO material. 

For the Si (890~$\Upomega$), the unthinned CCO $\parallel$ (430~$\Upomega$) and especially for the metallic Mo device (186~$\Upomega$), which are rather thick compared to standard TEM specimen thicknesses, the contact resistance is a large contribution to the measured total resistance. In case of the Mo device, a reliable measurement of the conductivity is thus not possible, which would require a strong decrease of the sample size. For the unthinned CCO $\parallel$ device, the high conductivity ($\sigma$~= 5.8~10\textsuperscript{3}~Sm\textsuperscript{-1}) is explained by the contribution from the Pt cover. Finally, for the Si device, the calculated resistivity of 1.3~10\textsuperscript{-3}~$\Upomega$m is considerably lower than the indicated resistivity of the employed Si wafer (0.1~$\Upomega$m). This is caused by Ga implantation during FIB milling (approximately 0.5~at\%, EDX analysis in SI5), which acts as additional p-doping of the Si, and possibly as well by remaining Pt on the device surface.

Finally, the resistance of the MLC NT, which possesses a diameter of 100~nm and a length of 2.8~$\upmu$m, is 10~k$\Upomega$ and the contact resistance is only a small fraction of the total resistance. The derived conductivity amounts to 3.7~10\textsuperscript{4}~Sm\textsuperscript{-1}, a value which agrees with previous measurements \cite{Hettler.2020,Hettler.2024_prep}. This confirms that using the recently developed transfer process for nanomaterials \cite{Hettler.2024_prep} allows to study them in a pristine state. 

\subsection{Thermoelectric properties}

All the studied devices clearly show the induction of a thermovoltage, which increases in magnitude with the temperature gradient created along the material by the differential heating device. The experiments thus prove the possibility to measure thermoelectric properties of materials with the proposed setup. However, for a quantitative analysis, the temperature gradient needs to be known in dependence of the applied heating current $I_H$. We performed COMSOL simulations of the temperature distribution under vacuum within the silicon nitride membrane for different $I_H$ and specimen parameters. Figure~\ref{F:Fig4}c and f depict two temperature maps simulated for $I_H$~=~5~mA and assuming a small thermal conductivity (1 Wm\textsuperscript{-1}K\textsuperscript{-1}) of the specimen, on the order of magnitude of the CCO material. The maps illustrate the effect of one important device parameter: the length of the hole milled in the membrane between the contact pads. A short hole, similar in size to the devices analyzed so far, barely affects the temperature distribution in the silicon nitride membrane (left part in Figure~\ref{F:Fig4}e), leading to a small temperature difference between the pads. Only a long hole across the entire membrane forms an effective heat barrier and yields a higher temperature (gradient) for the same $I_H$, which then merely depends on the thermal conductivity and the geometry of the specimen. 

To test this simulation result, we milled a longer hole in the membrane of the CCO $\parallel$ (66~$\upmu$m) and the Mo device (77~$\upmu$m) and repeated the measurement of the voltage offset in dependence of $I_H$. SI10 shows the SEM images of the devices with long hole. The result is plotted in Figure~\ref{F:Fig4}d in comparison with the same devices with a short hole and the CCO $\perp$ device, for which a long hole (65~$\upmu$m) was milled directly upon preparation. The measured voltage offset strongly increases by almost two orders of magnitude for the CCO $\parallel$ device, while the sign changes for the Mo device, which only shows a small change in magnitude. This striking difference is explained by two aspects: the contribution of the Pt contacts and the large difference in thermal conductivity between the two devices. With regard to the first aspect, the contribution of the Pt contacts has been ignored so far. Pt has an absolute Seebeck coefficient (-5~$\upmu$VK\textsuperscript{-1}) with the same magnitude as Mo (5~$\upmu$VK\textsuperscript{-1}), but with inverted sign. In the device with a short hole, the temperature gradient along both contact paths is largely similar and they cancel each other out almost entirely and the contribution from the Mo device is dominant. In contrast, the temperature gradient along the Pt contacts is asymmetric and their contribution outweighs the contribution of the Mo device for a long hole. Regarding the second aspect, the thermal conductivity is much higher for the metallic Mo (139~Wm\textsuperscript{-1}K\textsuperscript{-1}) than for the CCO (3~Wm\textsuperscript{-1}K\textsuperscript{-1}) \cite{Torres.2022}. Therefore, the temperature gradient along the CCO $\parallel$ device strongly increases, leading to the observed strong increase in the voltage offset.

The voltage offset shows an approximate, but not exactly parabolic dependency on $I_H$. The deposited power in the heating element is given by $R_H\cdot I_H^2$, but $R_H$ increases with the temperature, adding an additional contribution to the otherwise parabolic relationship. Assuming a linear increase of $R_H$ with increasing temperature, a valid assumption for Pt in the studied temperature range, the temperature of the heating device $T_H$ is linearly related to the increase of the heating resistance with respect to its resistance at ambient temperature: $T_H$ $\propto$ $\Delta R = R_H(T_H)-R_H(T_0)$. Indeed, when plotting the voltage offset over $\Delta R$, a linear dependency is observed (Figure~\ref{F:Fig4}e). This indicates that the induced temperature gradient is as well linearly dependent on the temperature of the heating device due to the linear relation between heat flux and temperature gradient. Furthermore, the Seebeck coefficient of the material does not significantly change with temperature in this temperature range, which agrees with bulk studies \cite{Torres.2022}.

The COMSOL simulations indicate that only a hole milled across the whole membrane provides a full prevention of heat flux along the membrane of the chip. However, when opening the hole over the whole membrane, the mechanical tension created along the investigated material that links both sides of the membrane is too large, causing the fracture of the device (SI11).

To get a rough quantitative estimation of the Seebeck coefficient of the CCO specimens, a value for the temperature gradient needs to be obtained. Therefore, the heater temperature can be calculated from the measured resistance and the calibrated linear temperature coefficient of Pt (SI12), which results in a temperature of approximately 125 and 140\degree C for the highest applied heating currents for the CCO devices (Figure~\ref{F:Fig4}e). Although the COMSOL simulations don't show a quantitative agreement with this estimated temperate, they were used to estimate the ratio between heater temperature and the actual temperature gradient across the device to 3:1 (SI13). Under these assumptions, the Seebeck coefficients of the CCO $\perp$ and the CCO $\parallel$ devices are calculated to 135 and 125~$\upmu$VK\textsuperscript{-1}. Although the values correspond well with bulk values \cite{Torres.2022}, they have to be seen as a very rough estimation and the real values could differ considerably.  The difference between CCO $\perp$ and $\parallel$ devices could indicate a variation of the Seebeck coefficient with the lattice direction for the CCO material and a possible influence of grain boundaries. Further studies with a fully quantitative setup (see section~\ref{S:quant}) are required to provide more insights in this relationship and to determine the role of specific types of grain boundaries.

\subsection{Dynamic thermoelectric characterization during \textit{in-situ} crystallization of an amorphous Ge thin film}
\label{S:Dyn}

The experiments and analysis presented so far were performed in a static way without inducing structural changes in the specimen during the \textit{in-situ} TEM studies. Such a static analysis can provide important information on thermoelectricity as it allows to correlate the thermoelectric properties with the structure and composition of the entire studied specimen at high resolution down to the atomic scale. For example, by preparing devices with specific microstructures or defects, it would be possible to determine the contribution of a specific grain boundary on the Seebeck coefficient. However, an \textit{in-situ} approach typically refers to the ability to track a dynamic evolution of a specimen within the microscope. Such a dynamic evolution could be generated by increasing the temperature using the heating element, but the temperatures reachable with the used chip design are not high enough to induce significant changes in the analyzed specimens. Another approach is to employ an electrical current to induce a structural modification in the specimen mainly by Joule heating \cite{Hettler.2021}.

\medskip
\newpage

\begin{figure}[h]
    \centering
    \includegraphics[width=0.76\linewidth]{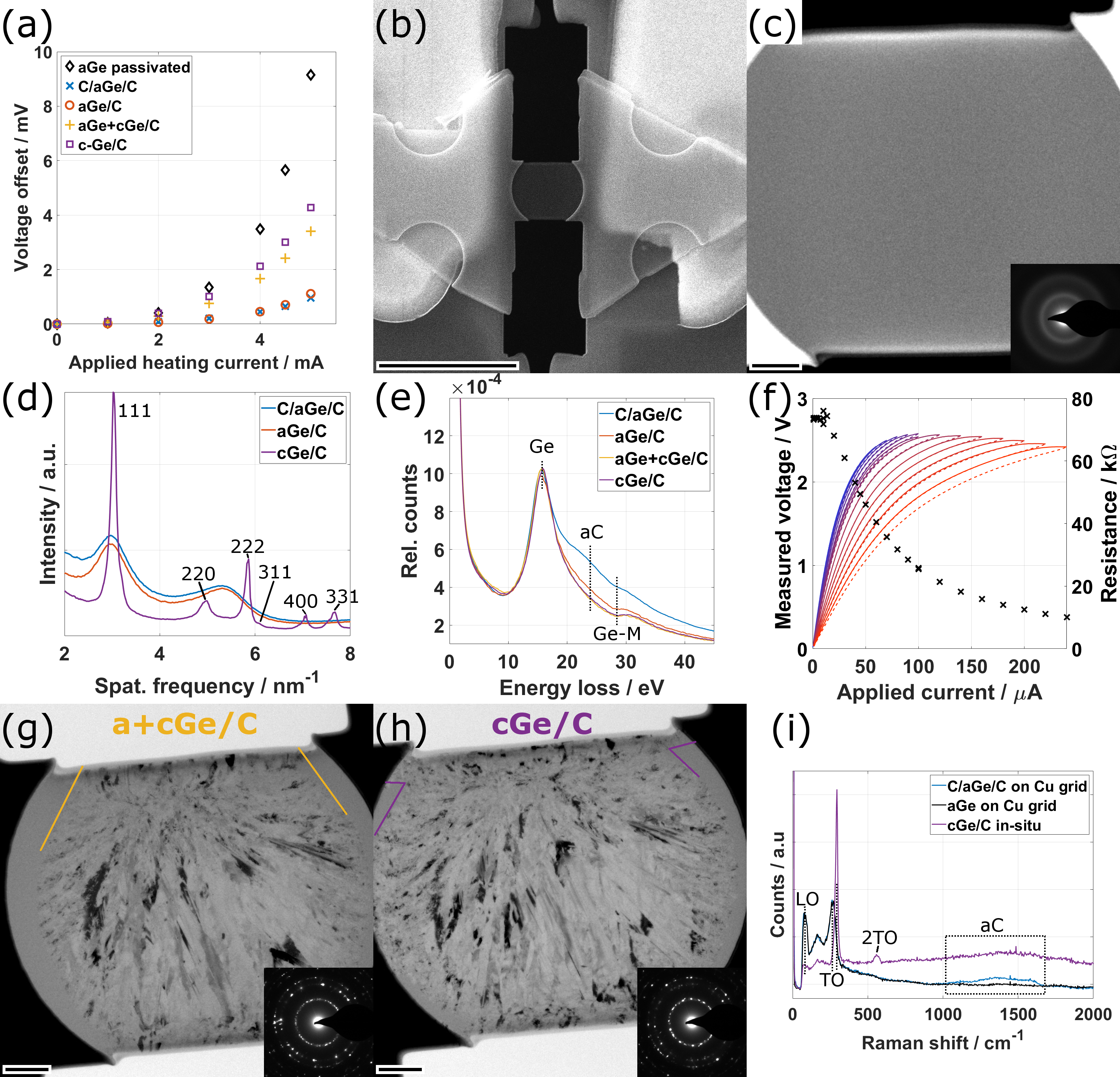}
    \caption{Dynamic \textit{in-situ} thermoelectric characterization of aGe-based thin films. (a) Voltage offset over applied heating current $I_H$ for a passivated aGe thin film (black diamonds), the aC-sandwiched aGe thin film as-prepared (blue crosses), after 1 min of plasma cleaning (red circles), after initial crystallization (yellow pluses) and after further crystallization (purple squares). (b) SEM image and (c) HAADF STEM image with corresponding SAED pattern of the as-prepared device. (d) Radial profiles of SAED patterns shown in (c,g,h) acquired at different steps throughout the experiment. Reflections corresponding to fcc Ge are indexed. (e) Low-loss EEL spectra acquired at different steps reveal the bulk plasmons of Ge and aC as well as the Ge-M edge. (f) Measured voltage over applied current for double-sweeps depicted by solid (up) and dashed (down) lines with color going from blue to red with increasing current. Calculated resistance at maximum applied current is shown by black crosses on the secondary y-axis for each sweep. (g,h) BFTEM images and corresponding SAED patterns of the specimen after (g) 1.5 min of plasma cleaning and (h) application of current sweeps shown in (f) reveal the crystallization of the specimen. (i) Comparison of Raman spectra from the \textit{in-situ} specimen after the experiment (purple) with as-prepared pristine aGe (black) and C/aGe/C (blue) thin films suspended on Cu grids. Scale bars are (b) 5~$\upmu$m, (c,g,h) 300~nm and the inset SAED patterns have a width of 14~nm\textsuperscript{-1}.}
    \label{F:Fig5}
\end{figure} 
\medskip
\newpage

As an example for such a dynamic \textit{in-situ} thermoelectric characterization, we tracked the induced thermovoltage during the crystallization process of an amorphous Ge (aGe) thin film upon application of elevated current densities. Therefore, a first \textit{in-situ} device with long hole (85~$\upmu$m) was prepared from an aGe thin film with a thickness of 36~nm (SI14). Its thermoelectric characterization is depicted by black diamond symbols in Figure~\ref{F:Fig5}a and reveals a high positive induced voltage offset of almost 10~mV at $I_H$~=~5~mA, even higher than the one measured for the CCO materials (Figure~\ref{F:Fig4}d). However, Ge rapidly forms a passivating oxidation layer after being exposed to air that generates an electrically insulating barrier (SI14) and impedes a controlled flow of the high electrical currents required for inducing an \textit{in-situ} crystallization of the thin film.

 In order to prevent this passivation, an aGe thin film was sandwiched between two amorphous C (aC) layers in a single evaporation step without venting the chamber. Figure~\ref{F:Fig5}b and c show a SEM and HAADF STEM image of the prepared device and a SAED pattern (inset in Figure~\ref{F:Fig5}c) confirms the amorphous nature of both aC and aGe layers. A radial profile of the SAED pattern is shown in Figure~\ref{F:Fig5}d by a blue line revealing two broad peaks centered around 3 and 5.4~nm\textsuperscript{-1}. A low-loss EEL spectrum (blue line in Figure~\ref{F:Fig5}e) confirms the presence of both Ge, reflected by the bulk plasmon at 16 eV and the Ge-M edge,\cite{Egerton.2011} and aC, which is visible from the bulk plasmon at 25 eV \cite{Hettler.2021}. A thermoelectric characterization of the device, which was prepared directly with a long hole (66~$\upmu$m), is depicted by blue crosses in Figure~\ref{F:Fig5}a and reveals a positive induced voltage with a magnitude ten times smaller compared to the pristine, passivated aGe thin film.

To reduce a possible impact of the aC layers on the measurement, an O\textsubscript{2}/Ar plasma cleaning step of 1~min was performed to remove most of the carbon from the specimen. The effect of the cleaning is reflected in a slight decrease of scattered intensity in the SAED pattern (red line in Figure~\ref{F:Fig5}d) and by an almost complete removal of the aC bulk plasmon in the low-loss EEL spectrum (red line in Figure~\ref{F:Fig5}e) when compared to the initial state (blue lines). A measurement of the voltage offset showed a slight increase of the induced voltage offset (red circles in Figure~\ref{F:Fig5}a), despite the increase in device resistance from 42~k$\Upomega$ to 64~k$\Upomega$ due to the removal of a part of the aC layer. As spectroscopic techniques indicated the presence of remaining C in the specimen (SI15), a second plasma cleaning step of 30s was performed. Indeed, the low-loss EEL spectrum reveals a further decrease in intensity in the region of the aC bulk plasmon (yellow line in Figure~\ref{F:Fig5}e), although spectroscopic analysis showed still a considerable presence of C (SI15). A large portion of the aGe thin film crystallized during this step, probably induced by an accidental electrical discharge. This crystallization into nano-grains is seen from the BFTEM image and the inset SAED pattern in Figure~\ref{F:Fig5}g. The border between crystallized and remaining amorphous areas is indicated by yellow lines. This crystallization translated to a more than three-fold increase of the induced voltage offset (yellow pluses in Figure~\ref{F:Fig5}a).

Following this unintended initial crystallization, we applied a series of double current sweeps with increasing maximum current up to 240~$\upmu$A to continue the crystallization process in a controlled way. The sweeps are shown in Figure~\ref{F:Fig5}f, where solid and dashed lines depict the up- and down-sweeps, respectively, and the color changes from blue to red with increasing maximum current. The calculated resistance at the maximum current of each sweep is plotted by black crosses and is observed to be constant (73~k$\Upomega$) up to an applied current of 13~$\upmu$A before starting to decrease due to the induced Joule heating and a related increase of conductivity of the semiconducting Ge. Above an applied current of approximately 70~$\upmu$A, the up- and down-sweeps start to differ considerably, indicating a structural modification of the specimen induced by the electrical current. This modification is linked to, firstly, a growth of the total size of the crystallized area and, secondly, a growth of the grain size in the central area. The first trend is observed from the corresponding BFTEM image (Figure~\ref{F:Fig5}h), which reveals that almost the entire suspended thin film has crystallized and only small amorphous regions remain, as indicated by purple lines. The second aspect is visible from the decreasing number of reflection spots in the SAED pattern (inset in Figure~\ref{F:Fig5}h), which correspond to fcc Ge (Crystallography Open Database \#7101739) as marked in the purple radial profile in Figure~\ref{F:Fig5}d. The crystal structure is as well confirmed by HAADF STEM imaging (SI15). The dynamic increment in crystallized area can be observed from the series of BFTEM images in supplementary movie S1 taken after each of the double current sweeps. This controlled increase in crystallization leads to a decrease in (static) resistance down to 18~k$\Upomega$, which goes in hand with a further rise of the induced voltage offset (purple squares in Figure~\ref{F:Fig5}a). 

The crystallization of the specimen is clearly seen by the comparison of Raman spectra acquired from the device after the experiment and from as-prepared thin films suspended on standard Cu TEM grids (Figure~\ref{F:Fig5}i). While the aGe specimens (blue and black line in Figure~\ref{F:Fig5}i) show broad peaks at 85 and 265~cm\textsuperscript{-1} corresponding to the longitudinal (LO) and transversal optical (TO) modes, a sharp TO mode, shifted to 295~cm\textsuperscript{-1}, is the dominant contribution in the spectrum of the crystallized \textit{in-situ} specimen (purple line), which also exhibits the 2TO mode. These features correspond well to Ge reference spectra from literature \cite{Zanatta.2020}. A contribution of the aC is seen by a very broad peak in the range between 1000 and 1700~cm\textsuperscript{-1} for the sandwiched aGe specimen and in strongly reduced intensity for the \textit{in-situ} specimen.

This \textit{in-situ} experiment demonstrates the possibilities of performing dynamic thermoelectric characterizations combined with TEM characterization, which allows the correlation of the microstructure, crystallinity and composition of a specimen at high spatial resolution with its thermoelectric properties. The experiment also can serve to illustrate that these measurements can be conducted with minimum and negligible contamination if the recently developed transfer method is employed \cite{Hettler.2024_prep}. In EDX spectra acquired from the aGe specimens, no Pt contamination could be found and Ga implantation was limited to the very edge of the specimen (SI15). The EDX analysis indicates a percentage of less than 0.5 at\% of Ga and less than 0.1 at\% of Pt in the central part of the thin film. A better estimation of a maximum amount of implantation is given by the measured electrical resistances assuming Ga as p-doping element in the Ge thin film. Neglecting the contribution from the remaining aC layer and the edges, the final resistance of 18 k$\Upomega$ leads to a resistivity of 0.34~$\Omega$mm, which gives an upper estimate of Ga concentration of 30~ppm, following the relation in \cite{Cuttriss.1961}. The concentration is expected to be even lower, due to the contribution from the aC and the fact that nano-sized objects typically exhibit higher conductivities compared to bulk values.

A comparison of the measured voltage offsets $V_0$ allows the following conclusions. Generally, the positive value for $V_0$ agrees with the positive sign of the Seebeck coefficient of Ge \cite{Hui.1967}. The magnitude of the measured $V_0$ for the crystallized aGe device is comparable to the CCO devices (Seebeck coefficient of approximately 150~$\upmu$VK\textsuperscript{-1}, Figure~\ref{F:Fig4}d), which agrees with the expected magnitude for Ge from literature \cite{Hui.1967}. The fact that the removal of a big part of the aC layer from the sandwiched aGe  specimen only leads to a minimum increase of $V_0$ (blue crosses and red circles in Figure~\ref{F:Fig5}a) suggests that the Seebeck coefficient is dominated by the aGe even if it is encapsulated by aC, which has a significantly higher electrical conductivity but a negligible Seebeck coefficient. This could be attributed to the amorphous nature of the entire specimen, which could cause a strong intermixing of charge carriers between the layers of the different materials.  The small increase in $V_0$ could be (partially) caused by an increased temperature gradient along the specimen due to a reduced thermal conductivity after removal of the aC layer.

$V_0$ is observed to increase with the degree of crystallinity, which is attributed to the increase in electrical conductivity of the Ge layer, allowing for a higher mobility of the charge carriers due to a reduced number of sources of electron scattering within the thin film. This finding as well agrees with literature \cite{Hui.1967}. Further \textit{in-situ} TEM studies could shed light on the dependence of the Seebeck coefficient on specific crystal orientations and/or the number and type of grain boundaries. Finally, the highest $V_0$ is observed for the passivated aGe thin film, which shows that the insulating passivation layer has a strong impact on the Seebeck coefficient of the device. A possible explanation could be that the insulating GeO\textsubscript{x} surface layer on its own contributes to the measured thermovoltage. Another contribution could be from the reduced thermal conductivity of the passivated aGe compared to the sandwiched specimen, which would cause a difference in temperature gradient.

\subsection{Toward fully quantitative thermoelectric characterization}
\label{S:quant}

\begin{figure}[t]
    \centering
    \includegraphics[width=0.6\linewidth]{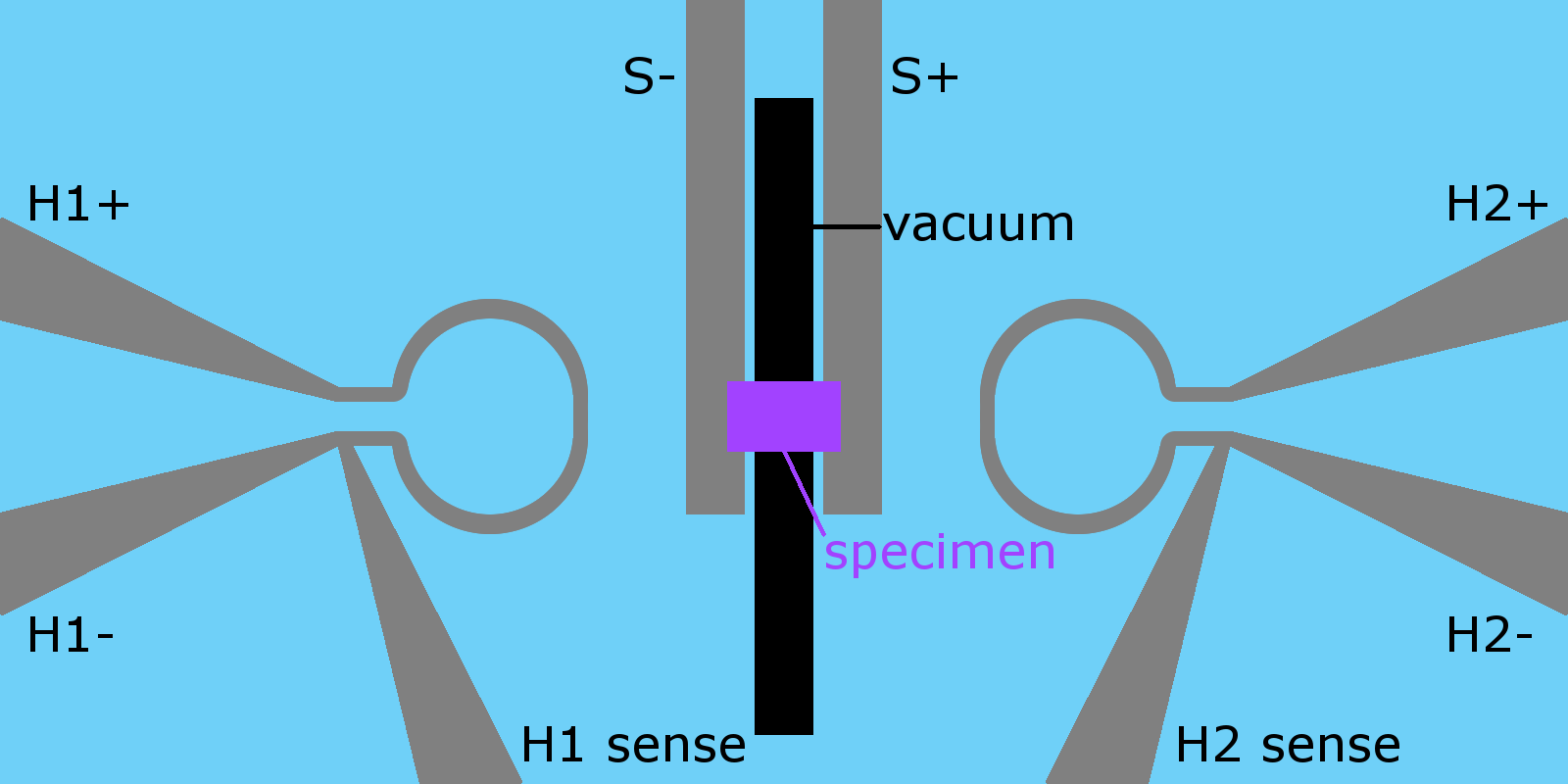}
    \caption{Sketch of the proposed improved chip design for thermoelectric \textit{in-situ} TEM characterization using a holder with eight electrical contacts as detailed in the text. The shown part should be placed on top of a membrane to facilitate the FIB milling of the hole between the contacts. }
    \label{F:Fig6}
\end{figure}

Despite the strong induced thermovoltage, its dynamic increment with increasing crystallization of an aGe thin film and the good agreement of the estimated Seebeck coefficients for the CCO materials, a reliable quantification still requires an exact knowledge of the temperature gradient between the contact pads, which cannot be determined from the employed chip without assuming a value for the thermal conductivity and calculations/simulations of the heat transfer of the device. Due to this limitation, the used chip design is not suitable to fully exploit the advantage of performing thermoelectric measurements by \textit{in-situ} TEM, which would allow studying the interplay between material properties and structure/composition on the local scale.

To reach the possibility of a full quantification, Figure~\ref{F:Fig6} shows a sketch of a device design, which could facilitate such measurements. The design would require an \textit{in-situ} TEM holder with at least eight electrical contacts. While two contacts would be used for two-probe electrical characterization of the device (S-, S+), the remaining six would be designated to two fully symmetrical heating elements. For both, two are used as current supply (H1,2-, H1,2+) and the third one as temperature sensor (H1,2 sense). Temperature measurement is performed in a three-point setup requiring identical dimensions of the two current paths for the heating elements. 

The chip design proposed in Figure~\ref{F:Fig6} considers the following requirements:

\begin{itemize}
    \item Fully symmetrical layout aiming for a homogeneous temperature distribution and a minimization of potential mechanical tension along the investigated material due to a stress difference between the two sides of the separated membrane.
    \item Strongly localized heat generation reducing the required length of the hole to form an effective conductivity barrier.
    \item Possibility to simultaneously control both the sample temperature and the temperature gradient.
\end{itemize}

The design shown in Figure~\ref{F:Fig6} would have to be placed on top of a silicon nitride membrane. With such a design, it would be possible to determine the temperature gradient and thus to quantify the Seebeck coefficient as well as the conductivity of the studied specimens. For example, for the CCO material it would be possible to disentangle the contributions from crystal orientation, grain boundaries or even pores. The design would also allow to perform a controlled \textit{in-situ} heating by setting both heating elements to the same temperature and thus follow, e.g., the thermoelectric properties of a material during annealing.  

\section{Conclusion}
To conclude, our experiments prove that a thermoelectric characterization of micro- and nanosized devices by \textit{in-situ} transmission electron microscopy (TEM) is possible. A thermovoltage is clearly induced in all studied devices and the sign corresponds to the sign of the materials' Seebeck coefficient. The results indicate that the temperature gradient is linearly proportional to the deposited heat energy and that an effective heat barrier should be generated between hot and cold side of the device to obtain a defined temperature gradient along the device. We propose an improved chip design and measurement setup for \textit{in-situ} TEM holders with eight electrical contacts, which should provide the combination of full quantitative thermoelectric characterization, including thermal and electrical conductivities,  with TEM and electron spectroscopies.  The detailed analysis of the devices also shows that, especially for nanomaterials, specimens can be prepared with negligible artifacts allowing their investigation in a pristine state. 

Such studies will allow to investigate in detail, e.g., the effect of  crystal directions in asymmetric materials,  grain boundaries and dopants, in principle down to the single-atom level, on the thermoelectric properties of bulk and nanomaterials. The experiments can be conducted in a static  or dynamic way. Static characterization is yielded by preparing various devices with specific structural properties such as exhibiting a certain type of grain boundary or a determined crystal direction with respect to the heat and current flow as exemplary shown for two CCO devices. Dynamical studies allow the tracking of thermoelectrical properties  during, e.g., \textit{in-situ} annealing of thermoelectric materials, as exemplary demonstrated by the \textit{in-situ} crystallization of an amorphous Ge thin film by Joule heating. The \textit{in-situ} studies can be further enriched by additional, complementary microscopy techniques performed on the identical device, such as Raman microscopy.

The \textit{in-situ} TEM technique lends itself on the one hand for fundamental studies as the entire device can be analyzed in depth and thus be well compared to first-principles calculations that take into account the defects present in the specimen. On the other hand, the evolution of the properties of a thermoelectric material during annealing allows a more applied approach of the technique. 

\medskip
\section*{Supplementary material}
Supplementary material is available under the following link: \href{}{https://doi.org/10.1016/j.ultramic.2024.114071} 

\medskip
\section*{Acknowledgements}
The authors acknowledge funding from the European Union’s Horizon 2020 research and innovation programme under the Marie Sklodowska-Curie grant agreement No 889546, the Government of Aragon (DGA) through the project E13\_23R, the Spanish MICIU with funding from European Union Next Generation EU (PRTR-C17.I1) promoted by the Government of Aragon and by the Spanish MICINN \newline (PID2019-104739GB-100/AEI/10.13039/501100011033                             and CEX2023-001286-S MICIU/AEI /10.13039/501100011033                           ) . The microscopy works have been conducted in the Laboratorio de Microscopias Avanzadas (LMA) at Universidad de Zaragoza. Sample courtesy (MLC NT) from MB Sreedhara and R. Tenne (MLC NT, Weizmann Institute of Science, Israel) is acknowledged. The authors thank M. Rengifo (INMA, CSIC-Universidad de Zaragoza) for discussions about electrical characterizations, P. Strichovanek (INMA, CSIC-Universidad de Zaragoza) for Ti-Pt sputter deposition, R. Valero (LMA, Universidad de Zaragoza) for support with MEMS and Ge thin-film fabrication and N. Navascues (INMA, CSIC-Universidad de Zaragoza) for support with Raman acquisition.
\medskip

\bibliographystyle{citstyle.bst}
\bibliography{TEbib}

\end{document}